# Snapshot light-field laryngoscope

**Shuaishuai Zhu,**[a,b] **Peng Jin,**[b,*] **Rongguang Liang,**[c] **Liang Gao**[a,d,*]

[a] University of Illinois at Urbana-Champaign, Department of Electrical and Computer Engineering, 306 N. Wright St., Urbana, USA, 61801

[b] Harbin Institute of Technology, Center of Ultra-precision Optoelectronic Instrument, 2 Yikuang St., Harbin, China, 150080

[c] University of Arizona, College of Optical Sciences, Tucson, USA, 85721

[d] University of Illinois at Urbana-Champaign, Beckman Institute for Advanced Science and Technology, 405 N. Mathews Ave., Urbana, USA, 61801,

**Abstract**. The convergence of recent advances in optical fabrication and digital processing yields a new generation of imaging technology—light-field cameras, which bridge the realms of applied mathematics, optics, and high-performance computing. Herein for the first time, we introduce the paradigm of light-field imaging into laryngoscopy. The resultant probe can image the three-dimensional (3D) shape of vocal folds within a single camera exposure. Furthermore, to improve the spatial resolution, we developed an image fusion algorithm, providing a simple solution to a long-standing problem in light-field imaging.

**\*Fourth Author**, E-mail: gaol@illinois.edu;   **\*Second Author**, E-mail: p.jin@hit.edu.cn

## 1 Introduction

Currently, approximately 7.5 million people in the United States suffer from voice disorders due to either trauma or diseases. Human vocal fold vibration is a complex three-dimensional (3D) movement. An unusual 3D shape of the vocal fold is a hallmark of a variety of vocal diseases, such as polyps, nodules, recurrent nerve paralysis, and cancer[1,2]. The acquisition of 3D data can facilitate the theoretical modeling of vocal fold dynamics, providing insights into vocal fold pathology[3,4] and fundamental phonation research[5,6].

The standard in-office methods for diagnosing voice disorders include videostroboscopy[7] and high-speed videoendoscopy[8]. Both techniques image only the horizontal movement of vocal folds. They cannot measure the movement of vocal folds along the air flow direction. Despite its vital importance, 3D laryngeal imaging is currently only available via a few methods— namely, computed tomography (CT)[3,9], magnetic resonant imaging (MRI)[10,11], laser triangulation[12,13], and optical coherence tomography (OCT)[14,15]. Although CT, MRI, and OCT can measure the full 3D profile of vocal cords, the prolonged acquisition time restricts their use in dynamic imaging. In addition, CT and MRI are costly, and they require special operating rooms. Alternatively, laser triangulation features a high acquisition speed. Nonetheless, it measures depths at only selected points or lines, resulting in a limited field of view. The lack of an en face image jeopardizes the sensitivity and specificity of diagnosis.

To enable fast imaging of vocal folds in 3D, for the first time, we introduce the paradigm of light-field imaging[16] into laryngoscopy. The resultant system, which we term a light-field laryngoscope (LFL), can capture a volumetric image of vocal folds within a single snapshot. Rather than acquiring only two-dimensional ($x$, $y$) ($x$, $y$, spatial coordinates) images, light-field cameras acquire both the spatial and angular information of remittance. The resultant four-dimensional (4D) ($x$, $y$, $\theta$, $\phi$) ($\theta$, $\phi$, 2D light emitting angles) datacube can be mathematically converted into a 3D ($x$, $y$, $z$) ($z$, depth) image through postprocessing[17]. Since no scanning is required, the 3D frame rate is limited by only the camera's data readout bandwidth. Although the light-field imaging was first proposed by Lippmann[18] in 1908, not until the last decade were breakthroughs achieved in demonstrating its biomedical applications. For example, Noah et al. developed a light-field otoscope for 3D imaging of the tympanic membrane in vivo[19]. Amir et al. constructed a light-field endoscope using a hexagonal liquid crystal lens array[20]. Massino et al. showed the potential of light-field cameras in retinal imaging[21]. Lastly, using a light-field microscope, Robert et al. recorded neuronal activity in 3D with an unprecedented frame rate[22].

## 2  Light-field laryngoscope design

We show the optical schematic and a photograph of the distal end of LFL probe in Fig. 1(a) and 1(b), respectively. The illumination light is guided to the tip of the probe through a multimode glass fiber (Thorlabs M28L01) and reflected towards the object by a right-angle prism (Edmund 84-506). The back-reflected light is collected by an objective lens (Edmund 49-657, $f = 18$ mm), forming an intermediate image $S_1$ at the distal end of a gradient-index (GRIN) lens (Gradient Lens Corporation COAT14-45-219; length, 219 mm (one pitch)). This intermediate image is then relayed by the GRIN lens to its proximal end, followed by being magnified by an optical system which consists of a microscope objective (Nikon CF Plan, $f = 40$ mm, $NA = 0.13$) and a tube lens (Thorlabs AC254-100-A, $f = 100$ mm). The magnified image is directed towards two imaging channels by a beamsplitter. While the transmitted image is directly measured by a high-resolution detector array (Point Grey CR-POE-20S2C-CS), the reflected image is acquired by a custom light-field camera which comprises a microlens array (MLA) (Advanced Microoptic Systems GmbH APO-Q-P148-R0.73, $f = 1.6$ mm) and a detector array (Point Grey BFLY-PGE-20E4M-CS). Herein we adopt a 2.0 light-field camera configuration—the distance between the MLA and the detector array is smaller than the focal length of the MLA[16]. We summarize the imaging parameters of two channels above (high-resolution (HR) and light-field (LF)) in Table 1. The geometrical dimensions of our prototype probe are similar to those of commercial laryngoscopes, with a nominal working distance of 65 mm and an outside diameter around 10 mm.

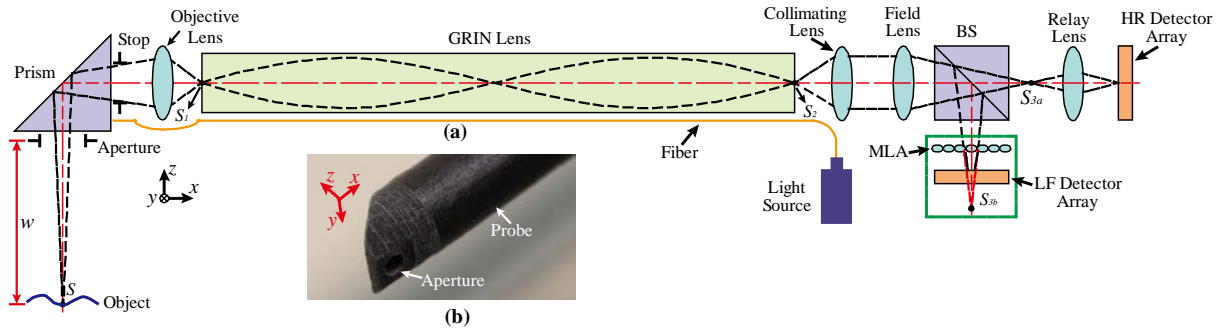

**Fig. 1** Schematic of a light-field laryngoscope. (a) Optical setup. The detector arrays in the HR and LF channel are referred to as HR and LF detector array, respectively. (b) Photograph of the distal end of the probe. GRIN, Graded-index; BS, Beam Splitter; MLA, Microlens Array; HR, High-Resolution; LF, Light-field.

Table 1 Specifications of two channels.

| | Spatial Resolution (pixels) | Depth Precision | Depth Range |
|---|---|---|---|
| High-resolution channel | 1920×1080 | Not applicable | Not applicable |
| Light-field channel | 640×360 | 0.37 mm | 62.5 mm – 67.5 mm |

Figure 2 shows the image processing pipeline, which consists of four steps, namely I) resolution enhancement, II) disparity estimation, III) depth reconstruction, and IV) combination of depth map and HR image. In Step I, we first derive the resolution ratio between the HR image and a single elemental image by imaging a calibration object (a chess board). Then we super-resolve each elemental image in the LF image using the HR image as the reference by a patch-based image super-resolution algorithm[23]. We downsample the HR image by a factor of the resolution ratio, followed by extracting a series of image patch pairs $\{h_i, l_i\}_{i=1}^{n}$. Here $h_i$ and $l_i$ denote the image patches extracted from the original and the downsampled HR images, respectively, and $i$ is the index enumerating the image patches. We save these image patches in dictionary $D_{ref}$. For each patch $p_j$ (5×5 pixels) in an elemental image, we search in $D_{ref}$ and identify nine patches $\{l'_k\}_{k=1}^{9}$ which have the smallest distances in the $L_2$ norm from $p_j$. We estimate the high-resolution representation $\hat{h}_j$ of $p_j$ by

$$\hat{h}_j = \frac{\sum_{k=1}^{9} w_k h'_k}{\sum_{k=1}^{9} w_k}, \qquad (1)$$

where $w_k = exp \frac{-\|p_j - l'_k\|^2}{2\sigma^2}$. Here $\sigma^2$ is a hyper-parameter, and we determine its value by using the Stanford light-field database[24] as a cross validation dataset.

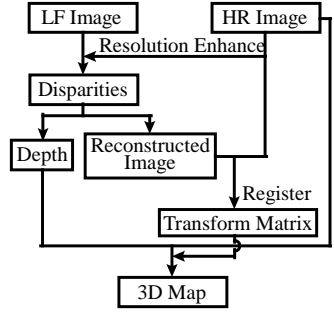

**Fig. 2** Flowchart of the image processing pipeline.

In step II, we consider the MLA as an array of stereo cameras and derive disparities from the correspondent elemental images pairwise. We illustrate the underlying principle using a simplified one-dimensional example (Fig. 3). The purple dashed lines denote the chief rays associated with microlenses. The extensions of these light rays (red dashed lines) converge to a virtual image point $S_{3b}$. Figure 3(b) shows the elemental images $M_1$, $M_2$, and $M_3$ formed behind the correspondent microlenses. For each elemental image pair, we identify the matched features by a searching algorithm based on correlation distance[25]. In brief, we first extract feature sets $\{f_{i,1}\}_{i=1}^{m_1}$ and $\{f_{i,2}\}_{i=1}^{m_2}$ from two elemental images, respectively. For each feature $f_{i,1}$, we search the correspondent neighborhood in $\{f_{i,2}\}_{i=1}^{m_2}$ and identify $f'_{j,2}$ which has the smallest correlation distance to $f_{i,1}$. We term $f'_{j,2}$ as the matched feature of $f_{i,1}$. Next, we calculate the disparity, $D$, as the relative distance between these two matched image features (Fig. 3(b)).

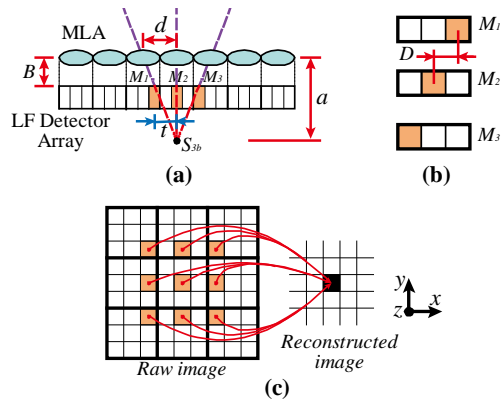

**Fig. 3** Light-field reconstruction. (a) Image formation in one dimension. (b) Zoomed-in view of elemental images $M_1$, $M_2$, and $M_3$. The disparity $D$ is calculated as the relative distance between two matched image pixels. (c) Two-dimensional image reconstruction.

In Step III, we derive depths through disparities. As shown in Fig. 3(a), in the global coordinate, we use $t$ to denote the absolute distance between the two matched pixels in $M_1$ and $M_2$. Then the disparity can be calculated by $D = d - t$, where $d$ is the MLA pitch. Using trigonometric relations, we get

$$(a - B)/a = t/d, \qquad (2)$$

where $B$ is the distance from the MLA to the LF detector array, and $a$ is the distance from the MLA to the virtual image $S_{3b}$. Substituting $d$ with $d = D + t$ yields

$$a = B \times d/D. \qquad (3)$$

To calculate the object depth, $w$, we project the virtual image $S_{3b}$ back to the object space. The relation between $w$ and $a$ can be experimentally determined through calibration[26].

In Fig. 3(c), we further generalize the scheme above to the two-dimensional case. In each elemental image, the orange pixel collects the light rays converging to the same virtual intermediate image point, $S_{3b}$. We group these pixels and map their values to a single pixel in the intermediate image. Following this procedure pixelwise yields a reconstructed image[25].

In light-field imaging, there is a trade-off between the spatial and angular resolution because the total number of 4D light-field datacube voxels cannot exceed the total number of sensor pixels. To some extent, this trade-off can be mitigated by employing compressed sensing algorithms[27-29]. However, these techniques are computationally extensive, and they highly rely on the ill-posed assumption that the light-field is sparse in a given domain. Also, the requirement of multiple camera exposures[27,28] makes them unsuitable for imaging dynamic scenes. By contrast, in the proposed LFL, we alleviate this problem through fusing the depth map with a high-resolution reference image in Step IV. We warp the depth map to the actual size of HR image through a transform matrix derived by registering the reconstructed image from the LF channel and the HR image. Then we mathematically combine the warped depth map and the HR image to generate a high-resolution 3D representation of the original scene.

## 3 Experiments

The depth precision in the LFL is determined by a myriad of factors, namely pitch size of MLA, the distance from the MLA to the sensor, the pixel size the sensor, $NA$ and the vignetting[30]. To evaluate the depth precision in our prototype, we scanned a point source along both the $x$- and $z$-axes. Twelve steps were taken along the $x$-axis with step size set as 0.8 mm, while nine steps were taken along the $z$-axis with step size set as 0.64 mm. The mean and standard deviation of the measured depths along the $x$-axis at each $z$-axis step are shown in Fig. 4. The black dashed line shows the ground truth. The root mean square error (RMSE) of the average measured depth along the $x$-axis is 0.07 mm. The depth precision is estimated as the average standard deviation along the $z$-axis. The result approximates 0.37 mm, providing an effective depth-to-resolution greater than ten.

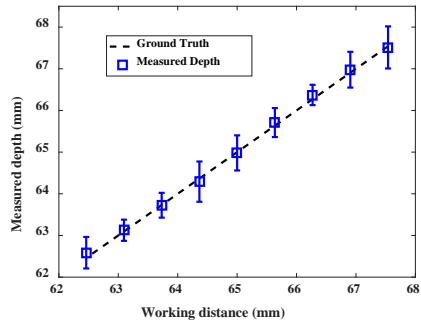

**Fig. 4** Quantification of depth precision in light-field laryngoscope.

To assess the lateral resolution, we imaged a 1951 USAF resolution test target at the nominal working distance (65 mm) of the LFL. Figure 5(a) shows the reconstructed image of the test target with a zoomed-in inset view of bars in Group 4 and 5. In Fig. 5(b), we plot the intensity profile along a green dashed line in Fig. 5(a) (inset). The Lord Rayleigh's criterion states that two overlapping slit images are resolvable when the irradiance of the saddle point between two fringes is lower than $8/\pi^2$ times of the maximum irradiance[31]. Based on this criterion, the bars of group 4 element 4, as shown in the orange dashed rectangle, are the finest resolvable features. Therefore, the lateral resolution of the LFL is 22.6 lp/mm.

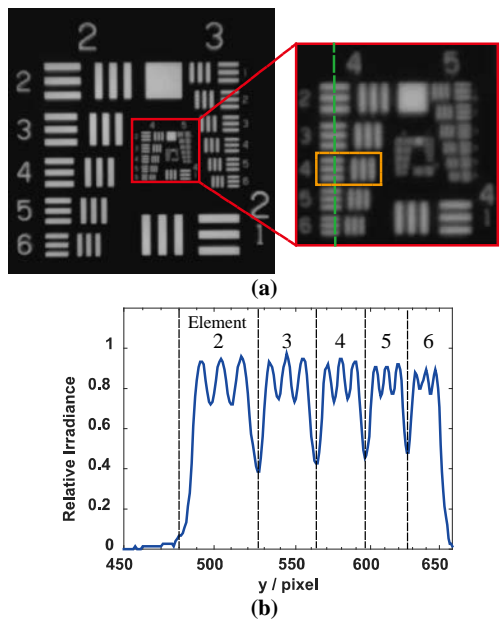

(a)

(b)

**Fig. 5** Spatial resolution of the light-field laryngoscope. (a) Reconstructed in-focus image of a 1951 USAF resolution target. (b) Intensity profile along the green dashed line in Fig. 5(a) inset.

Next, we compare this value to the diffraction limit, which is calculated as $R_d = 1/2r = NA/1.22\lambda$, where $r$ is the radius of airy disk, $\lambda$ is the wavelength of the incident light, and $NA$ is the numerical aperture. Given $\lambda = 0.6$ µm and $NA = 0.02$, we have $R_d = 27.3$ lp/mm, which is greater than the experimental resolution of the LFL. We attribute this discrepancy to two reasons. First, we constructed the LFL prototype using only off-the-shelf lenses. The cumulative geometric aberrations blur the image. Second, the employment of a GRIN lens introduces a considerable level of chromatic aberration[32], which also degrades the imaging performance. Although beyond the scope of this paper, we can potentially overcome these problems by replacing these off-the-shelf lenses and GRIN lens with custom ones.

Finally, to demonstrate the 3D imaging capability of the LFL, we performed two phantom experiments. First, we used a tilted paper surface with letters as an object. A reference photograph is shown in Fig. 6(a). Captured by a single snapshot, the reconstructed 3D image is shown in Fig. 6(b). The recovered surface tilt angle matches with the experimental setup. Next, we imaged a vocal fold phantom (Fig. 6(c)), using "vessels" and "vocal fold edges" as features for disparity estimation. Because LFL measures depths only at distinct feature points, we filled the blank areas using interpolation on the assumption that the phantom surface is naturally continuous in slope and curvature. Figure 6(d) shows the reconstructed 3D image, agreeing well with the ground truth.

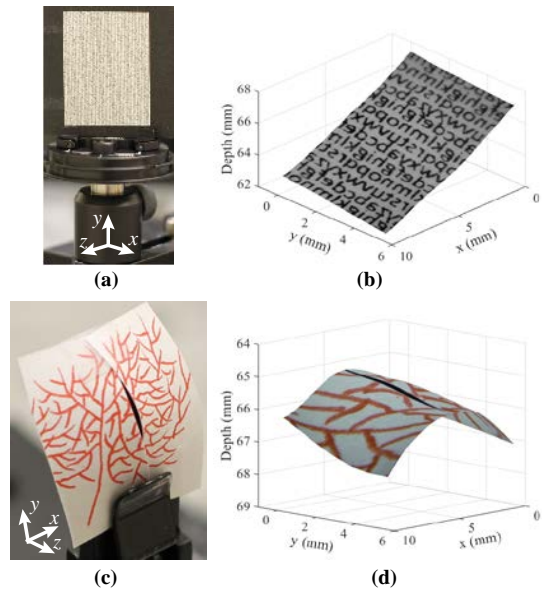

**Fig. 6** 3D phantom imaging. (a) Reference photograph of a tilted paper surface with letters. (b) Reconstructed 3D image. (c) Reference photograph of a vocal fold phantom. (d) Reconstructed 3D image.

The overall layout of LFL is simple, and it can be built using only off-the-shelf optical components. The projected manufacturing cost is comparable to the conventional medical laryngoscopes that are routinely used in the primary care clinics. Also, thanks to its simplicity, the system requires rudimentary training to operate.

## 4   Conclusions

In summary, we constructed a 3D imaging LFL. Rather than measuring only the spatial information, the LFL acquires the spatial and angular information of the incident light rays in parallel. Such a measurement leads to a recovery of a 3D representation of the original scene with high fidelity. Due to a snapshot acquisition format, the 3D imaging speed is limited by only the camera's readout speed, which can be potentially up to 1,000 volumes/sec when coupled to a high-speed camera[33]. In light of its unprecedented 3D imaging performance, we anticipate that LFL will open a new area of investigation in both the clinical diagnostics and fundamental phonation research.

*Acknowledgments*

This work was supported in part by NSF CAREER grant (1652150) and discretionary funds from UIUC. We thank Kuida Liu for his contribution to the super-resolution algorithm just as James Hutchinson for the close reading of the manuscript. We also gratefully acknowledge the financial support from the China Scholarship Council.

Biographies and photographs for the other authors are not available.

**Caption List**

**Fig. 1** Schematic of a light-field laryngoscope. (a) Optical setup. The detector arrays in the HR and LF channel are referred to as HR and LF detector array, respectively. (b) Photograph of the distal end of the probe.

**Fig. 2** Flowchart of the image processing pipeline.

**Fig. 3** Light-field reconstruction. (a) Image formation in one dimension. (b) Zoomed-in view of elemental images $M_1$, $M_2$, and $M_3$. The disparity $D$ is calculated as the relative distance between two matched image pixels. (c) Two-dimensional image reconstruction.

**Fig. 4** Quantification of depth precision in light-field laryngoscope.

**Fig. 5** Spatial resolution of the light-field laryngoscope. (a) Reconstructed in-focus image of a 1951 USAF resolution target. (b) Intensity profile along the green dashed line in Fig. 5(a) inset.

**Fig. 6** 3D phantom imaging. (a) Reference photograph of a tilted paper surface with letters. (b) Reconstructed 3D image. (c) Reference photograph of a vocal fold phantom. (d) Reconstructed 3D image.

**Table 1** Specifications of two channels.